\documentclass[10pt]{article}
\usepackage{graphicx}

\usepackage{color}

\usepackage{lineno}

\title{State-dependent fragmentation of protonated uracil and uridine}

\author{Martin Pitzer$^{1,2}$ $^\ast$, Christian Ozga$^{2}$, Catmarna K{\"u}stner-Wetekam$^{2}$, \\
Philipp Rei{\ss}$^{2}$, Andr{\'e} Knie$^{2}$, Arno Ehresmann$^{2}$, \\
Till Jahnke$^{3}$, Alexandre Giuliani$^{4}$, Laurent Nahon$^{4}$\\
\normalsize{$^{1}$Department of Chemical and Biological Physics,}\\
\normalsize{Weizmann Institute of Science,} \\
\normalsize{P.O. Box 26, 7610001 Rehovot, Israel}\\
\normalsize{$^{2}$Institute of Physics and Center for Interdisciplinary Nanostructure Science and Technology (CINSaT),}\\
\normalsize{University of Kassel,}\\
\normalsize{Heinrich-Plett-Stra{\ss}e 40, 34132 Kassel, Germany}\\
\normalsize{$^{3}$Institute for Nuclear Physics,}\\
\normalsize{Goethe-University,}\\
\normalsize{Max-von-Laue-Stra{\ss}e 1, 60438 Frankfurt, Germany}\\
\normalsize{$^{4}$Synchrotron Soleil,}\\
\normalsize{L{\'{}}Orme des Merisiers, Saint Aubin, 91192 Gif-sur-Yvette, France}\\
\\
\normalsize{$^\ast$To whom correspondence should be addressed; }\\
\normalsize{E-mail:  martin.pitzer@weizmann.ac.il} \\
}

\date{\today}

\begin{document}

\newcommand{\mpitzer}[1]{\textcolor{blue}{\emph{#1}}}
\newcommand{\ev}[1]{$ #1 \; \mathrm{eV}$}

\maketitle

\begin{abstract}
Tandem mass spectroscopy ($\textrm{MS}^2$) combined with single photon excitation in the VUV range (photon energy \ev{4.5-9}) was performed on protonated uracil ($\textrm{UraH}^{+}$) and uridine ($\textrm{UrdH}^{+}$). The precursor ions with $m/z\;113$ and $m/z\;245$ respectively were produced by an Electrospray Ionization source (ESI) and accumulated inside a quadrupole ion trap mass spectrometer. After irradiation with tunable synchrotron radiation, product ion mass spectra were obtained. Fragment yields as a function of exciting energy show several maxima that can be attributed to the photo-excitation into different electronic states. For uracil, vertically excited states were calculated using the equation-of-motion coupled cluster approach (EOM-CCSD) and compared to the observed maxima.
This allows to establish correlations between electronic states and resulting fragment masses and can thus help to disentangle the complex deexcitation and fragmentation pathways of nucleic acid building blocks above the first electronically excited state. Photofragmentation of the nucleoside uridine shows a significantly lower variety of fragments, indicating stabilization of the nucleobase by the attached sugar.
\end{abstract}
 \\

\section{Introduction}

Many experimental techniques have been used to investigate the interaction of electromagnetic radiation with building blocks of DNA (desoxyribonucleic acid) and RNA (ribonucleic acid). Understanding the response of these small molecular species to the absorption of an energetic photon (from near to vacuum ultra-violet) may provide insight into radiation damage of living tissue and may also have astrobiological implications regarding their stability in the interstellar medium. For better comparison with theoretical models, investigating isolated - i.e. solvent-free - molecules in the gas phase is often the method of choice.\\
Early works on nucleobases and related compounds include absorption measurements of neutral gas phase molecules with lamp-based spectrophotometers in a vapour cell \cite{Clark1965} or resonance-enhanced multiphoton ionization (REMPI) from a supersonic jet \cite{Brady1988}; the observed absorption maxima have been interpreted with the help of quantum chemical calculations \cite{Barbatti2010}. Photoelectron spectroscopy has revealed details of the electronic structure \cite{Dougherty1976}, by means of time-resolved \cite{Satzger2006} or highly energy-resolved \cite{Fulfer2015, Chen2016, Zhao2018} measurements. Mass spectroscopy has been used in various approaches to determine ionization energies of biologically relevant molecules and appearance energies of their fragments (e.g. \cite{Jochims2005}). For recent reviews on the interaction of neutral nucleic acid building blocks with UV and X-ray radiation, see \cite{Prince2015, Schwell2015}.\\
As an intermediate between neutral gas-phase molecules and species in solution, protonated and microhydrated molecules are often considered to better represent environmental conditions than their neutral counterparts \cite{Pandey2017}. 
Cation photodissociation processes provide an additional benefit for mass spectroscopic (MS) investigation: Even at low excitation energies, charged fragments are produced that can easily be detected and quantified by MS, contrary to the dissociation products of neutral molecules.
Several works have been based on quadrupole traps to select a charged precursor which is then photoexcited by a UV-laser. Protonated nucleobases have extensively studied with such set-ups, and in particular the role of microsolvation was investigated \cite{Pedersen2013, Pedersen2014}, as well as the effect of different tautomers \cite{Marian2005, Cheong2011, Berdakin2014, Matthews2017} and the excited state relaxation dynamics \cite{Broquier2017}.\\
However, by using near and middle UV photons available from these sources, only the first and the onset of the second singlet excited electronic state can be probed. As a consequence, only the break-ups with relatively low dissociation energy are visible. In this paper we present a complete study over the entire VUV range, showing the photodynamics of uracil and its nucleoside uridine up to 9 eV photon energy.

\section{Experimental}
\begin{figure}
\includegraphics[width = 15cm, keepaspectratio=true]{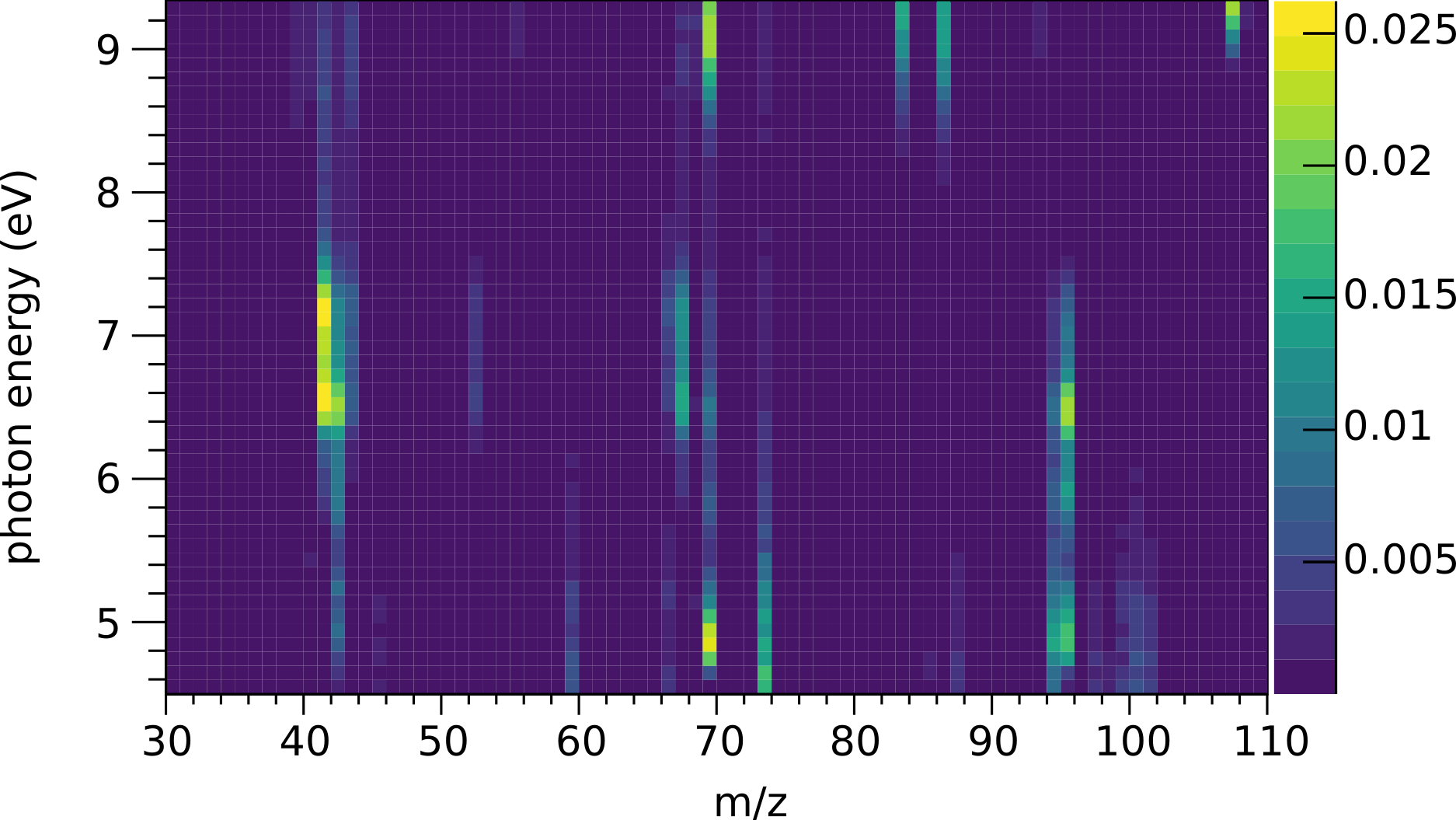}
\caption{Two-dimensional fragment map after photoexcitation of protonated uracil (precursor $m/z\; 113$, not shown because of high abundance). The fragment mass is plotted on the $x$-axis, the exciting photon energy on the $y$-axis. The color indicates the relative yield, normalized for the photon flux. Several masses with low appearance energies (\ev{5-7}) can be identified.}
\label{fig:uracil_map}
\end{figure}

In tandem mass spectroscopy, a specific precursor ion (cation or anion) is mass-selected and stored, usually in a quadrupole ion trap. Basically all excitation mechanisms (photons, electrons, ions, neutral atoms) can be used to activate the stored ions and thus induce fragmentation. After the interaction time is over, the resulting fragment mass spectrum is read out.\\
The present experiments were performed with the SRMS2 setup \cite{Milosavljevic2012} on the DESIRS beamline \cite{Nahon2012} at Soleil synchrotron facility (France), using photons of variable energy for activation - an approach sometimes designated as action spectroscopy.\\
The protonated species were created upon electrospray ionization. For uracil, $0.3\; \textrm{mM}$ aqueous solution, and for uridine, $1\; \textrm{mM}$ aqueous solution were used. The precursor ions were accumulated for $50\; \textrm{ms}$ in the trap and then irradiated with linearly polarized photons for $5\; \textrm{s}$. In order to maximize the photon flux ($ \sim 10^{13}\; \textrm{photons}/s$), we used wide monochromator slits corresponding to a bandwidth of typically $9\;\textrm{meV}$. Several photon energy scans were done in $100\;\textrm{meV}$ steps; at each photon energy, mass spectra were recorded for four minutes, yielding 45 fragment spectra per photon energy that were averaged. All spectra were normalized for the photon flux that was measured with a photo-diode prior to the experiments. To avoid ionization and fragmentation by higher orders of the synchrotron radiation, a suprasil window was used in the lower energy range (\ev{4.5-7.1}), and a krypton gas filter in the higher energy range (\ev{7-9.5}). This change in settings leads to differences in photon flux for the two scan regions. The relative scaling of the measured fragment yields between these two photon energy regions was found by requiring consistent values at the overlapping photon energies (\ev{7.0-7.1}). Below \ev{5}, the photon flux drops significantly so that only the strongest fragmentation channels yield meaningful results; above \ev{9.5}, ionization of neutral molecules from the residual gas or from contaminants in the source leads to a background which prevents a quantitative analysis. We therefore limited our study to the \ev{4.5-9.5} range.\\
The main source of uncertainty in the present measurements are fluctuations between individual mass spectra recorded at the same photon energy, i.e. statistical variations in particle numbers due to the low abundance (``shot noise''). We therefore use the standard deviation of the fragment yield for the errorbars. It can be seen that the uncertainty is higher for low photon energy where the fragment yield is low, and that it is comparable for different masses at a given photon energy. As a consequence, only errorbars for one fragment species are plotted in the spectra for better visibility. Systematic uncertainties such as the measurement of the photon flux are considered negligible compared to the statistical error.\\
For an overview, the resulting data can be displayed in a two-dimensional fragment map (Figure \ref{fig:uracil_map}), with the fragment mass-to-charge ratio $m/z$ on the $x$-axis and the exciting photon energy on the $y$-axis. This representation allows to identify interesting fragment masses, while plotting the yield of selected fragments as a function of photon energy enables a more quantitative spectroscopic analysis.

\section{Computational Details}
\begin{figure}
\includegraphics[width = 15cm, keepaspectratio=true]{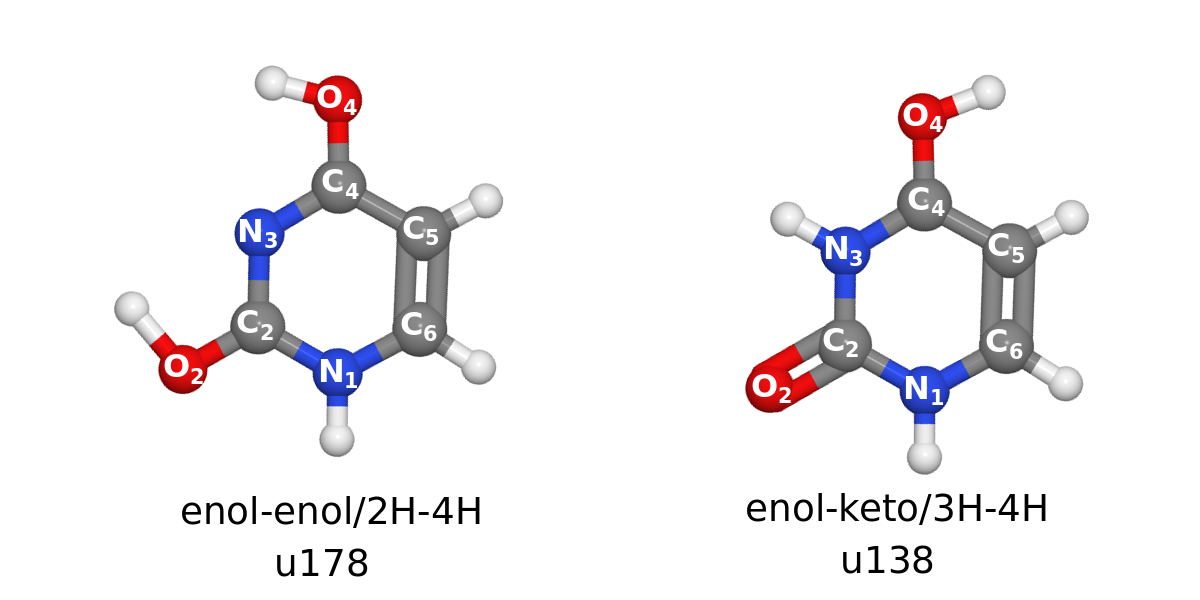}
\caption{The two energetically lowest tautomers of protonated uracil. The enol\textendash enol/2H\textendash 4H form (u178) is considered the ground state, while the enol\textendash keto/3H\textendash 4H (u138) contributes 1 to $20\;{\%}$ of the population at room temperature.}
\label{fig:uracil_tautomers}
\end{figure}

Calculation of excited state energies was performed with GAMESS(US) \cite{Schmidt1993}. So far, only few theoretical works are available that deal with electronically excited states of nucleobases in the measured energy range.
Epifanovsky et al. \cite{Epifanovsky2008} studied in detail the effect of the basis set and the choice of the excitation algorithm for neutral uracil. Before calculating the excitation energies of protonated uracil (which has the same number of electrons and - as is generally assumed for protonated nucleobases \cite{Broquier2017,Sadr-Arani2014, Pedersen2014}- the same planar $C_{s}$ symmetry) we benchmarked our numerical settings against those results.\\
For geometry optimization of the ground state, density functional theory (DFT) with the B3LYP functional \cite{Becke1993,Stephens1994,Hertwig1997} and the 6-311++G(2d,2p) basis set \cite{Krishnan1980} was used. Vertical excitation energies were calculated with EOM-CCSD (equation-of-motion coupled cluster with single and double excitation) \cite{Piecuch2002, Kowalski2004, Wloch2005} and the 6-311+G(d,p) basis set. These parameters were found to yield reliable results for the first few states (both for singlet and triplet states). States with larger Rydberg state contribution ($n \geq 3$) are very sensitive to the basis set and the level of theory \cite{Epifanovsky2008} are therefore omitted here. Excited state optimization or vibrational analysis was beyond the scope of this work due to the expected complex shape of the potential energy landscape \cite{Marian2005}.\\
Protonated uracil exists in several tautomers \cite{Berdakin2014, Pedersen2014, Sadr-Arani2014}. In our calculation, we consider the two structures that are depicted in Figure \ref{fig:uracil_tautomers}. They have been consistently identified as the two energetically lowest tautomers in the cited works. Estimates of the enthalpy difference and the relative population vary strongly. The higher-lying tautomer u138 (enol\textendash keto form) is considered to contribute between 1 and $20\;{\%}$ to the population at room temperature \cite{Pedersen2014}.

\section{Results and discussion}
\subsection{Uracil}
\begin{figure}
\includegraphics[width = 15cm, keepaspectratio=true]{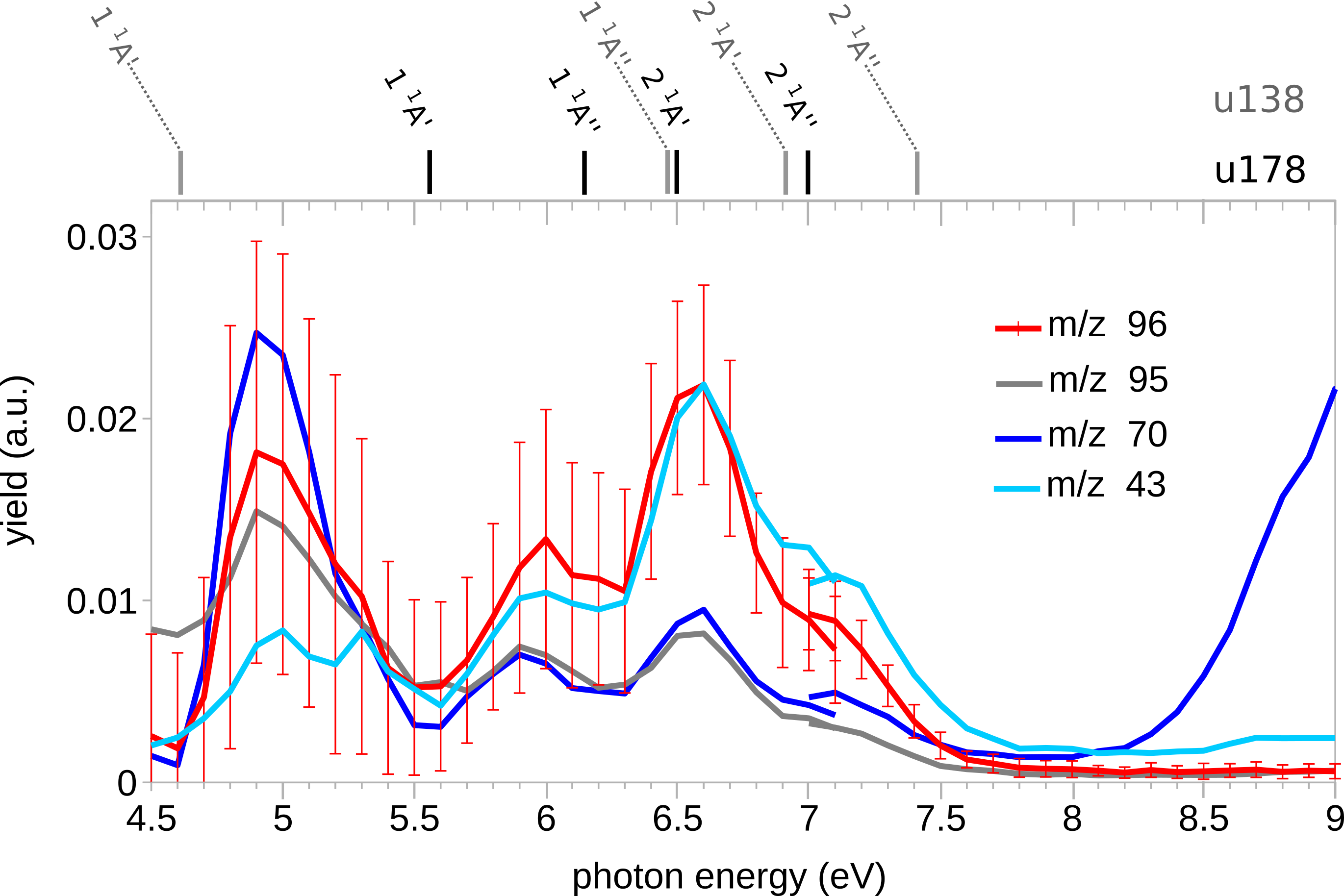}
\caption{Fragment yields of protonated uracil as a function of exciting photon energy. Only fragments that exhibit a significant yield at low energy (around \ev{5}) are displayed. The error bars represent the standard deviation of the individual mass spectra distribution, showcased for the $m/z\; 96$ fragment (see Experimental Section); the overlapping values at \ev{7} are due to a change in beamline settings. The vertical bars at the top indicate the vertical excitation energies calculated with EOM-CCSD for the two tautomers. The observed fragments are ascribed as follows: $m/z\; 96$: $\textrm{NH}_{3}$ loss; $m/z\; 95$: $\textrm{H}_{2}\textrm{O}$ loss; $m/z\; 70$: $\textrm{HNCO}$ loss; $m/z\; 43$: $\textrm{HNCO}^{+}$}
\label{fig:uracil_lowE}
\end{figure}

Figure \ref{fig:uracil_map} gives an overview which fragments of protonated uracil ($\textrm{C}_4\textrm{H}_5\textrm{N}_2\textrm{O}_2^+$) occur at which excitation energy. The fragment mass-to-charge ratio $m/z$ is plotted on the $x$-axis, the exciting photon energy on the $y$-axis. 
On the lower end of the energy spectrum, several peaks around \ev{5} and a broader region from 6 to \ev{7.5} can be discerned.\\
Figure \ref{fig:uracil_lowE} displays the yield of four fragments with low appearance energy. The repeated measurements at \ev{7} mark the overlap of two energy scans with different settings (see Experimental section). The fragment masses shown here are usually considered as $\textrm{NH}_{3}$ loss ($m/z\; 96$), $\textrm{H}_{2}\textrm{O}$ loss ($m/z\; 95$), $\textrm{HNCO}$ loss ($m/z\; 70$) and the cation $\textrm{HNCO}^{+}$ \cite{Nelson1994, Berdakin2014, Pedersen2014, Sadr-Arani2014}.\\
The vertical bars at the top of the figure indicate the calculated vertical excitation energies for the two considered tautomers. The calculated excitation energies and oscillator strengths with respect to the ground state are summarized in Table \ref{tab:excitation_energies}. For completeness, triplet state energies are shown in the table as well; since the electronic ground state has singlet character, transitions into these states are forbidden and omitted in the figures.
\begin{table}%
	\begin{tabular}[h!]{p{2cm}p{2cm}p{2cm}p{2cm}p{2cm}}
	\hline
	 & \multicolumn{2}{c}{tautomer u178} &  \multicolumn{2}{c}{tautomer u138}\\ \hline
	state & $\Delta E$ (eV) & $f$ & $\Delta E$ (eV) & $f$\\
	\hline
	1 ${}^{3}A'$  & 4.332 & forbidden & 3.695 & forbidden \\
	2 ${}^{3}A'$  &   5.060 & forbidden & 5.420 & forbidden \\     
	1 ${}^{1}A'$  &  5.564 & 0.096 & 4.607 & 0.108\\ 
	1 ${}^{3}A'' $ &   6.018 & forbidden & 5.994 & forbidden \\
	1 ${}^{1}A''$ &  6.169 & $8 \cdot 10^{-4}$ & 6.459 & $8 \cdot 10^{-6}$ \\  
	2 ${}^{1}A'$  &  6.498 & 0.069&  6.913  & 0.094 \\
	2 ${}^{1}A''$ &   6.996 & 0.005& 7.411 & 0.005\\ 
	 \\
	\end{tabular}
	\caption{Vertical excitation energies $\Delta E$ (singlet and triplet) and oscillator strengths $f$ (singlet) for excitation from the ground electronic state. Calculations were performed for the two energetically lowest tautomers (see Fig. \ref{fig:uracil_tautomers}), using EOM-CCSD with the 6-311+G(d,p) basis set.}
	\label{tab:excitation_energies}
	\end{table}
The strong peak around \ev{5} that has been observed previously \cite{Pedersen2014, Berdakin2014} is usually attributed to the first excited singlet state $S_1$ with the term symbol  $1 \; {}^{1}A'$. By cooling the protonated ions to cryogenic temperatures $(20 - 50\; \textrm{K})$, Berdakin et al. \cite{Berdakin2014} were able to disentangle the contributions of the two tautomers and the vibrational structure of this state. Using the empirical formula $E_{\mathrm{exp}} \approx E_{\mathrm{vert}}- 0.5\; \mathrm{eV}$, they attribute the region between 4.8 and \ev{5.4} to the first excited singlet state of the u178-tautomer, and a region between 3.9 and \ev{4.5} (not visible in our data) to the u138-tautomer. This rule accounts for the fact that no excited state optimization has been performed when calculating the vertical excitation energy. It is, however, contradictory to the observation that fragment yields can peak at higher energies than the vertical excitation energy due to activation barriers for the dissociation and a kinetic shift of the reaction \cite{Schwell2015}.\\
It should be noted that the lowest excited state of \textit{neutral} uracil is a so-called dark state, due to its ${}^{1}A''$ symmetry and the concomitant small oscillator strength for transition from the symmetric ground state. For both considered tautomers of \textit{protonated} uracil, this dark state is higher in energy than the bright $1\; {}^{1}A'$ state, so that the first maximum can be discussed without referring to complications such as intensity-borrowing by a lower-lying state \cite{Epifanovsky2008, Nachtigallova2011}.\\
In addition to this first peak, three partly overlapping peaks can be distinguished at 6.0, 6.6 and \ev{7.2}. Considering the small oscillator strength for the transition to the $1\;{}^{1}A''$ state, the most likely assignment for the first two peaks is $ 6.0\; \mathrm{eV} \Leftrightarrow 2\;{}^{1}A'\;(u178) $ and $ 6.6\;\mathrm{eV} \Leftrightarrow 2\;{}^{1}A'\;(u138) $. The third peak could be related to the $2\;{}^{1}A''$ of either tautomer, but additional higher-lying states could play a role as well.\\
Although all four masses follow the trend, differences are visible: While $m/z\; 70$, $m/z\; 95$ and $m/z\; 96$ show the strongest yield for the first observed excitation, $m/z\; 43$ is much more pronounced for higher energy. The large uncertainty at low photon energies, however, prevents a more quantitative analysis.\\

\begin{figure}
\includegraphics[width = 15cm, keepaspectratio=true]{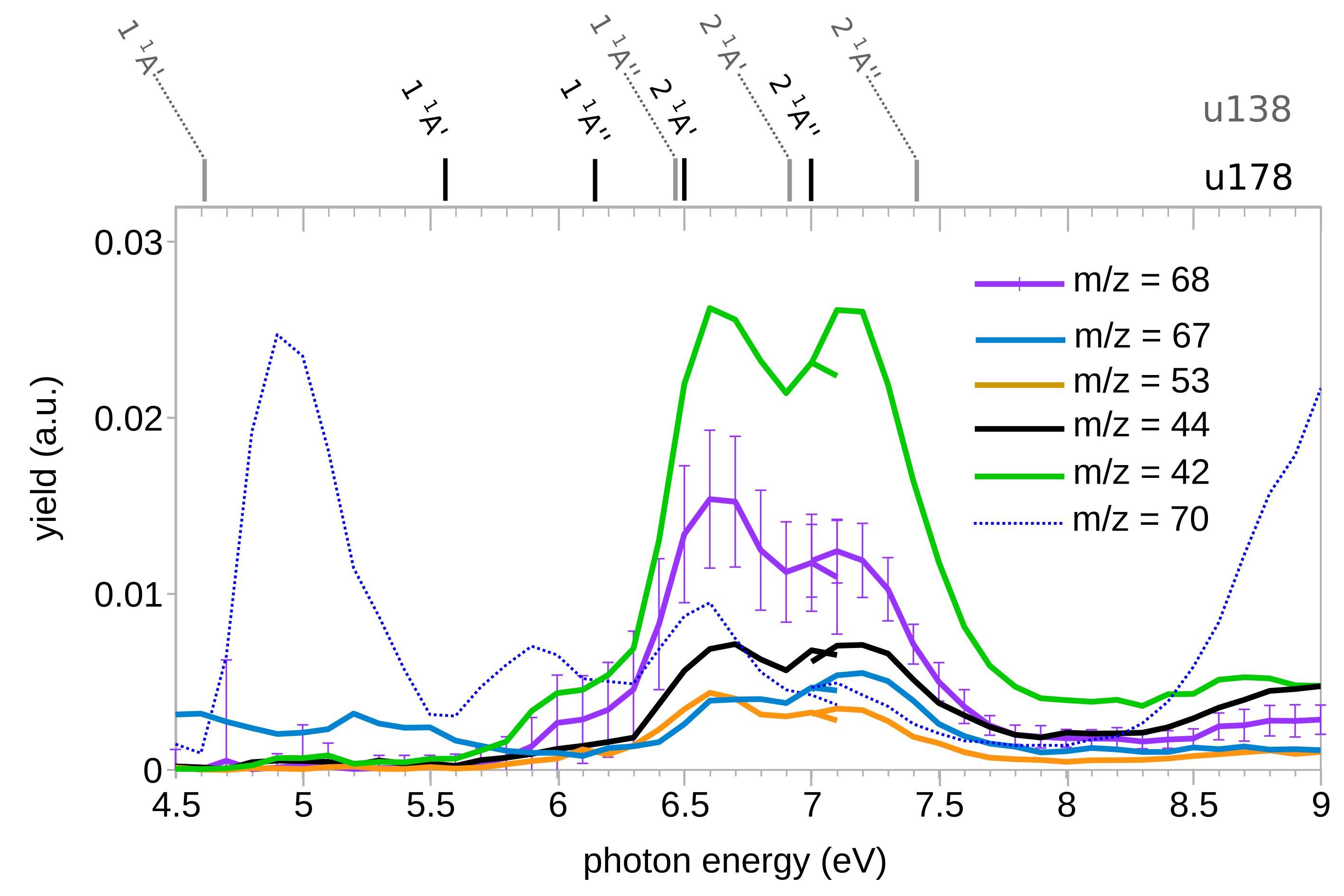}
\caption{Fragment yields of protonated uracil with appearance energies above \ev{5.5}, with errorbars indicating the standard deviation for the fragment $m/z\; 68$. The calculated vertical excitation energies are again drawn as grey bars. The fragment $m/z\; 70$ is sketched for better comparison with Figure \ref{fig:uracil_lowE}. Concerning the identification of the observed masses, see table \ref{tab:fragment_assignment}.}                                                                                                                                                                                                                                                          
\label{fig:uracil_highE}
\end{figure}

Figure \ref{fig:uracil_highE} shows additional fragments that have previously not been observed after photoexcitation of protonated uracil. They do not occur below \ev{5.5}, indicating that they cannot originate from the first electronically excited state. From DFT calculations, the fragment cation $m/z\; 42$ was attributed to $\textrm{H}_{2}\textrm{CCNH}_{2}$ \cite{Sadr-Arani2014} and $m/z\; 67$ as $\textrm{C}_{3}\textrm{N}_{2}\textrm{H}_{3}$; in that work, the two isomers $\textrm{HOCNH}$ and $\textrm{OCNH}_{2}$ are proposed for $m/z\; 44$ and various isomers of $\textrm{C}_{3}\textrm{NOH}_{2}$ for $m/z\; 68$. This chemical formulae are consistent with an early study using isotope-labelled species \cite{Nelson1994} that additionally identified $m/z\; 53$ as $\textrm{C}_3\textrm{OH}^+$. It should be noted that Sadr-Arani et al. \cite{Sadr-Arani2014} assumed partly different tautomers.\\
The previously discussed fragment $m/z\; 70$ is inserted to allow a better comparison with Figure \ref{fig:uracil_lowE}. In addition to the higher appearance energy, these fragments show slightly different features between 6 and 8 eV compared to the fragments in Figure \ref{fig:uracil_lowE}: Most strikingly, the peak at \ev{7.2} is as prominent or even higher than the one at  \ev{6.6} whereas for the fragments with low appearance energy, the latter is dominant.\\
Following the previous assignment, this means that the ``high-energy'' fragments predominantly decay from $2\;{}^{1}A''$ state. Since the calculations suggest a much higher transition probability into the $2\;{}^{1}A'$ state, we cannot exclude that high vibrational excitations play also a role in producing these fragments. In addition, it is remarkable that the yields of the ``high-energy'' fragments are almost proportional to each other.\\
Finally, we compare the grouping into ``high-energy'' and ``low-energy'' fragments to primary and secondary fragments as classified in a thorough study using various combinations of oxygen and carbon isotopes, as well as thio- and methyl-derivatives of uracil \cite{Nelson1994}: There, the cations with masses $m/z\; 96, 95, 70$ and possibly $44$ were attributed to a primary loss of neutral groups as discussed above. The cations $m/z\; 68,67,53,52,43,42$ and $28$ were identified as products from a secondary decay of the former fragments. Except for the $\textrm{HNCO}^{+}$ and $\textrm{H}_2\textrm{NCO}^{+}$ cations ($m/z\; 43,44$) and the unobserved fragments $m/z\; 52, 28$, the primary fragments are observed at lower photon energy whereas the secondary fragments only occur at higher excitation energies. This is remarkable since collision-induced dissociation is expected to primarily excite rovibrational states. Table \ref{tab:fragment_assignment} provides a summary of the assignment of observed masses to the chemical formulae and fragmentation pathways found in CID.\\
\begin{table}%
	\begin{tabular}[h!]{|p{1.5cm}p{4cm}p{2cm}p{2cm}p{2cm}|}
	\hline
	m/z & chem. formula (CID) & \multicolumn{3}{c|}{suggested decay mechanism (CID)} \\
	\hline
	96 &  $\textrm{C}_4\textrm{H}_2\textrm{N}_1\textrm{O}_2^+$ & $\textrm{C}_4\textrm{H}_5\textrm{N}_2\textrm{O}_2^+$ & - $\textrm{NH}_3$  & \\
	95 &  $\textrm{C}_4\textrm{H}_3\textrm{N}_2\textrm{O}^+$ & $\textrm{C}_4\textrm{H}_5\textrm{N}_2\textrm{O}_2^+$ & - $\textrm{H}_2\textrm{O}$ &  \\     
	70 & $\textrm{C}_3\textrm{H}_4\textrm{NO}^+$ & $\textrm{C}_4\textrm{H}_5\textrm{N}_2\textrm{O}_2^+$ & - $\textrm{HNCO}$ & \\ 
	43 & $\textrm{HNCO}^{+}$ & $\textrm{C}_4\textrm{H}_5\textrm{N}_2\textrm{O}_2^+$ & - $\textrm{HNCO}$ & - $\textrm{HCN}$  \\
	\hline
	68 &  $\textrm{C}_{3}\textrm{NOH}_{2}^+$ & $\textrm{C}_4\textrm{H}_5\textrm{N}_2\textrm{O}_2^+$ & - $\textrm{NH}_3$ & - $\textrm{CO}$  \\  
	67 &  $\textrm{C}_{3}\textrm{N}_{2}\textrm{H}_{3}^+$ & $\textrm{C}_4\textrm{H}_5\textrm{N}_2\textrm{O}_2^+$ & - $\textrm{H}_2\textrm{O}$ & - $\textrm{CO}$ \\
	53 & $\textrm{C}_3\textrm{OH}^+$  & $\textrm{C}_4\textrm{H}_5\textrm{N}_2\textrm{O}_2^+$ & - $\textrm{HNCO}$ & - $\textrm{NH}_3$\\ 
	44 & $\textrm{H}_2\textrm{NCO}^{+}$ & $\textrm{C}_4\textrm{H}_5\textrm{N}_2\textrm{O}_2^+$ & - $\textrm{C}_3\textrm{H}_3\textrm{NO}$ &\\
	42 & $\textrm{C}_2\textrm{NH}_{4}^{+}$ & $\textrm{C}_4\textrm{H}_5\textrm{N}_2\textrm{O}_2^+$ & - $\textrm{HNCO}$ & - $\textrm{CO}$ \\
	\hline
	\end{tabular}
	\caption{Identification of fragments and fragmentation pathways in collision-induced dissociation (CID) \cite{Sadr-Arani2014, Nelson1994} grouped according to appearance at low photon energy (top) or high photon energy (bottom) in the current study.}
	\label{tab:fragment_assignment}
	\end{table}

\subsection{Uridine}
\begin{figure}
\includegraphics[width = 15cm, keepaspectratio=true]{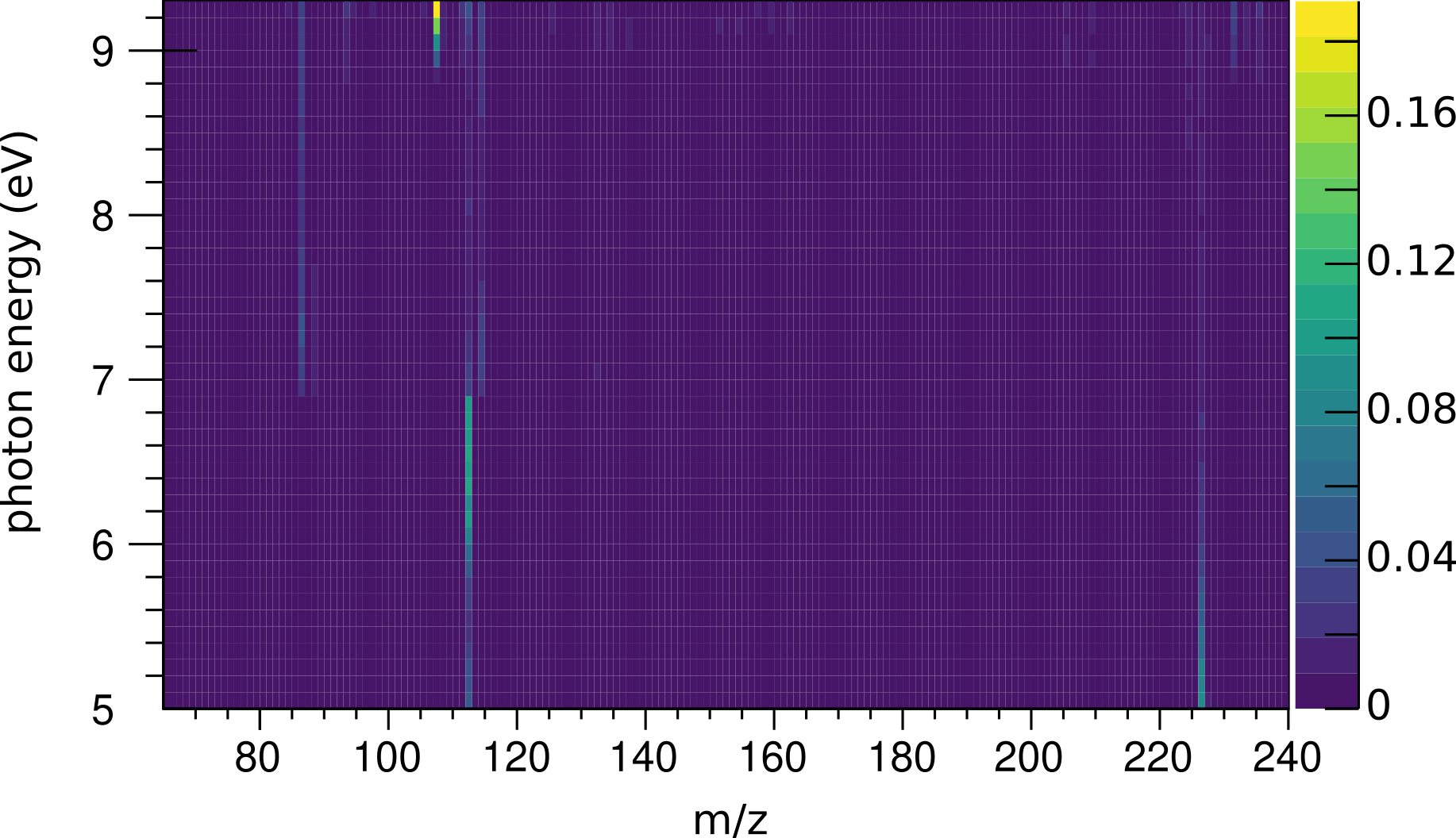}
\caption{Two-dimensional fragment map after photoexcitation of protonated uridine ($m/z\; 245$). Below \ev{9}, fragments from ring opening reactions are not present (cf. Figure \ref{fig:uracil_map}). The ions with $m/z\; 86$ and $m/z\; 115$ could not be assigned.}
\label{fig:uridine_map}
\end{figure}

\begin{figure}
\includegraphics[width = 15cm, keepaspectratio=true]{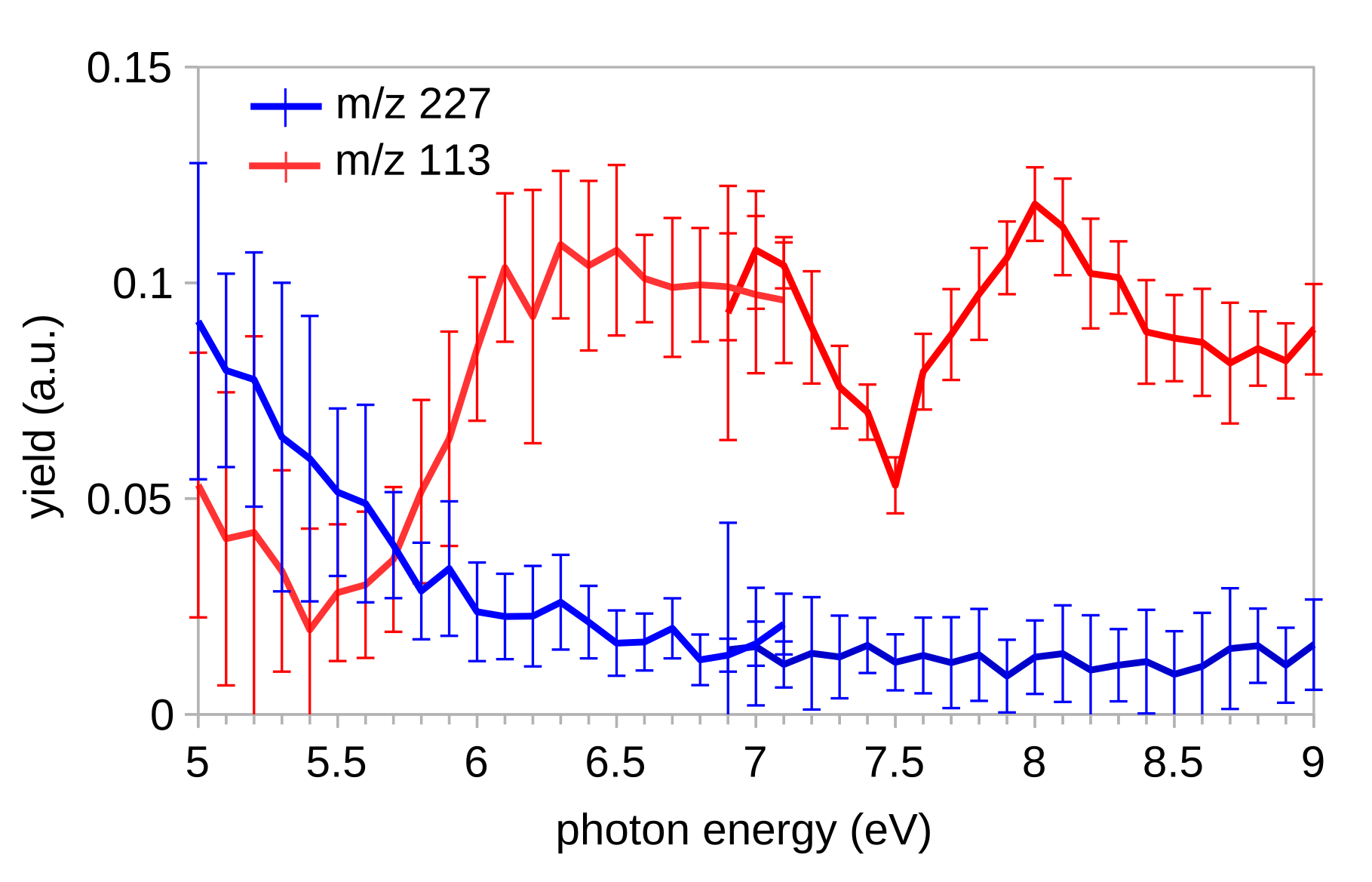}
\caption{Fragment yield for the two most abundant fragments of protonated uridine ($m/z\; 245$) as a function of exciting photon energy, with errorbars indicating the standard deviation of individual spectra. For low energies, water loss $m/z\; 227$ is the strongest decay channel. The yield of the protonated nucleobase UraH+ with $m/z\; 113$ follows the trend observed in the fragment spectra of UraH${}^{+}$.}
\label{fig:uridine_1D}
\end{figure}

Fragmentation of the nucleoside of uracil, uridine, was measured in the same exciting photon energy range (Fig. \ref{fig:uridine_map}). Due to instrumental limitations, masses below $m/z\; 65$ cannot be recorded for the uridine precursor $m/z\; 245$. Very few fragments are detected and dominant break-ups of protonated uracil such as $m/z\; 70$ or $m/z\; 95$ barely occur here. This can be clearly seen by comparing Figure \ref{fig:uridine_map} to the much richer fragment map of uracil in Figure \ref{fig:uracil_map}. This difference indicates stabilization of the nucleobase by the N-glycosidic bond. Below \ev{5}, the photon flux was too low to extract any meaningful fragment data. The main fragments for photon energies below \ev{9} are $m/z\; 227$ and $m/z\; 113$ that can be attributed to a water loss and the splitting into uracil and sugar moiety respectively. Similar fragment pattern have been observed in collision-induced dissociation of protonated uridine \cite{Dudley2014} and in photon-induced near-threshold ionisation of neutral uridine \cite{Levola2014, Pan2008}; the fragments  $m/z\; 86$  and  $m/z\; 115$ could not be assigned; it should be noted, however, that the former was also present in \cite{Levola2014}. The N-glysosidic bond cleavage has also been found to be the only decay channel in an infrared multiple photon dissociation (IRMPD) study of protonated uridine at low dissociation energies \cite{Wu2015}.\\
Figure \ref{fig:uridine_1D} shows in more detail the yield of these two fragments as a function of the photon energy. The splitting into nucleobase and sugar roughly follows the trend that was observed for the uracil fragments with peaks at 5, 6.5 and \ev{7.2}. This indicates that the corresponding electronic states of the nucleobase ``chromophore'' are excited, but decay differently than in the bare nucleobase. An additional peak occurs at \ev{8} where the uracil fragmentation shows a minimum. A meaningful calculation of excited electronic states of this molecule with 29 atoms and various conformers, was beyond our resources. However, it has previously been suspected that nucleosides internally convert into the electronic ground state via conical intersections, thus driving the excess energy into fragmentation of the weakest bond, the sugar-nucleobase link \cite{Nielsen2005}. Tautomeric changes due to the attached sugar might also play a role in stabilizing the nucleobase \cite{Wu2015}.\\

\section{Conclusion}
The fragment mass spectra of protonated uracil and protonated uridine have been recorded as a function of the exciting photon energy in the range of \ev{4.5-9.5}. This approach combines two attractive features: First, single photon excitation precisely defines the energy deposited in the system. 
Second, the additional charge present in the protonated form leads to charged fragments at lower excitation energy than for the neutral, making low-lying states accessible to mass spectroscopy.
The experimental data show that various fragment masses exhibit an energy-dependent yield with distinct maxima. Calculations allowed correlating some of these peaks to certain electronically excited states. In particular, fragments could be grouped into those that can already decay from the first singlet excited state ($m/z \in \{43,70,95,96\}$) and those originating from higher excitations ($m/z \in \{42,44,53,67,68\}$). This grouping corresponds very well to a distinction into primary and secondary fragments in collision-induced dissociation \cite{Nelson1994, Sadr-Arani2014}.\\
It should be kept in mind that the fragmentation process occurs on a complex potential landscape. Many groups have investigated in detail the relaxation dynamics of neutral uracil, e.g. \cite{Nachtigallova2011, Matsika2011, Zhou2012, Kleinermanns2013, Yu2016, Keefer2017, Ghafur2018}, and related species (for a recent review, see \cite{Boldissar2018}). These works illustrate that fast intersystem crossings to the electronic ground state can prevent fragmentation, and how these processes depend on isomerization and environment of the molecule. The present results only show the start and end point of this complex trajectory on the potential energy surfaces. Nevertheless, they can serve as benchmark for numerical investigation of the break-up mechanisms.\\
Energy dependent fragment mass spectra of uridine, the nucleoside of uracil, in the same photon energy range show a very different fragmentation pattern. The absence of low-mass fragments indicates that the sugar group provides deexcitation mechanisms that compete with ring opening reactions and thus stabilize the nucleobase itself, a finding which appears very relevant for the field of radiobiology.

\section{Acknowledgments}
We thank the whole SOLEIL staff for running smoothly the facility and for granting beamtime under proposal no. 20160870. For the computational work, we acknowledge support from the ChemFarm High Performance computing at Weizmann Institute, in particular from Dr. Mark Vilensky. This work was supported by the Agence Nationale de la Recherche (France), under project number ANR-08-BLAN-0065, and by the Deutsche Forschungsgemeinschaft (DFG) through Sonderforschungsbereich 1319 (ELCH).


\end{document}